# Home-Made Blues:

# Residential Crowding and Mental Health in Beijing, China

Xize Wang[a], Tao Liu[b, *]

**Abstract**

Although residential crowding has many well-being implications, its connection to mental health is yet to be widely examined. Using survey data from 1,613 residents in Beijing, China, we find that living in a crowded place—measured by both square meters per person and persons per bedroom—is significantly associated with a higher risk of depression. We test for the mechanisms of such associations and find that the residential crowding–depression link arises through increased living space-specific stress rather than increased life stress. We also identify the following subgroups that have relatively stronger residential crowding–depression associations: females, those living with children, those not living with parents, and those living in non-market housing units. Our findings show that inequality in living space among urban residents is not only an important social justice issue but also has health implications.

a. Department of Real Estate, National University of Singapore, Singapore. Email: wangxize316@gmail.com. OCRID: 0000-0002-4861-6002
b. College of Urban and Environmental Sciences and Center for Urban Future Research, Peking University. Email: wangxize316@gmail.com. OCRID: 0000-0002-3568-4882
* Corresponding author.

**Introduction**

Mental health issues have become an increasing global concern. For instance, one in every five adults in the United States has experienced mental illness at least once a year (Substance Abuse and Mental Health Services Administration, 2012), and one in every seven Singaporeans has experienced a mental disorder throughout their life (Institute of Mental Health, 2018). The prevalence of mental health problems has negative consequences for both individuals and societies. On the one hand, as identified in the United Nations Sustainable Development Goals (United Nations, 2015), mental health is an important aspect of people's well-being, and those with mental health issues normally have a lower health-related quality of life (Yang et al., 2018). On the other hand, mental health problems impose a high financial burden on society. Studies in China show that medical costs due to mental health issues are an estimated 1.1% of the national GDP, and the additional indirect cost (e.g., productivity loss) may be four times of the direct cost (Xu et al., 2016). Hence, improving the mental health status of urban dwellers would improve both their individual well-being and the fiscal sustainability of the public sector.

Urban living can negatively impact people's mental health status (Lederbogen et al., 2011). Among the myriad components of urban life, the urban physical environment has been shown to be a contributing factor in urban dwellers' mental health problems (Evans, 2003; Marmot, 2005). Specifically, the physical environment can increase individuals' stress levels, and those constantly exposed to stressors are more likely to have depression and other mental health issues (Turner et al., 1995). Thus, policy makers and urban planners can help to promote urban dwellers' mental health status by providing an enjoyable urban physical environment.

Housing, especially the provision of living spaces, is an important physical environment factor that policy makers can utilize to promote people's mental health. Existing empirical evidence has shown that living in a crowded place—whether measured by square meters or by number of rooms per person—is associated with poorer mental health (Foye, 2017; Hu & Coulter, 2017; Li & Liu, 2018). In Chinese cities, the marketization of housing has substantially increased the supply of living space to urban dwellers (Deng & Chen, 2019). However, although the *average living space* has greatly improved, the *distribution of living space* has become more skewed, as many urban dwellers, especially in big cities, have been left behind (Huang & Jiang, 2009). Studies have found that migrants are more likely to suffer from residential crowding and hence are more likely to experience mental health problems (Li & Liu, 2018; Xie, 2019). Likewise, local urban residents who were "relatively deprived" in China's real estate boom also merit attention (Y. Zhang & Chen, 2014).

The present paper aims to fill this gap by focusing on the residential crowding problem for long-term residents in China's big cities. Taking Beijing as a case study, we examine the relationship between residential crowding and depression in Beijing's *hukou* holders. We also explore the varying effect levels for different genders, household structures, and neighborhood types, aiming to identify subgroups that have higher "policy sensitivity" with respect to residential crowding. In addition, this study intends to contribute to the literature theoretically by testing whether residential crowding is associated with depression via increased life stress or increased living space-specific stress.

The reminder of the paper is organized as follows. The next section reviews the literature. The third section introduces the data source and analytical methods. The next two sections present and discuss the empirical findings, and the final section concludes.

**Literature Review**

The theoretical foundation of the relationship between residential crowding and mental health mainly lies in studies of the stress process (Pearlin & Bierman, 2013; Pearlin et al., 1981). Residential crowding, which normally involves either insufficient living space or lack of private space at home, lowers a person's sense of control over their surroundings (Evans et al., 2003; Rodin, 1976). The reduced sense of control is associated with high stress levels and a higher risk of depression (Pearlin et al., 1981; Steptoe et al., 2007). In addition to such residence-specific stress, residential crowding can also be associated with worse mental health status through increased life stress. Prior research has shown that people with limited living space have higher life stress due to reasons such as lower levels of social interaction (Evans et al., 2003; Foye, 2017). Residential crowding also can make people less "mentally resilient" against life stress. In other words, those living in a more crowded home will have a stronger association between life stress and depression, as they are less likely to heal from the cognitive fatigue brought about by life stress (Evans et al., 2003; Pearlin & Bierman, 2013). Figure 1 illustrates these three possible mechanisms whereby residential crowding may be connected to depression: (a) the "crowding-related stress" hypothesis, (b) the "mediation through life stress" hypothesis, and (c) the "moderating life stress" hypothesis.

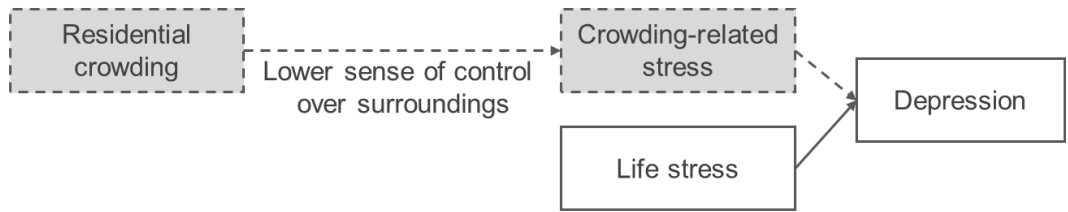

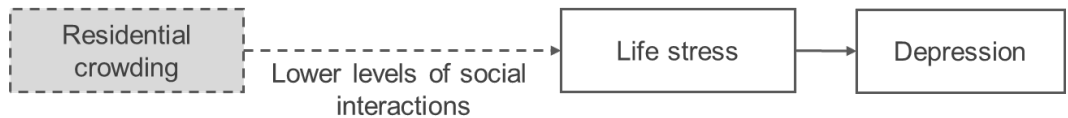

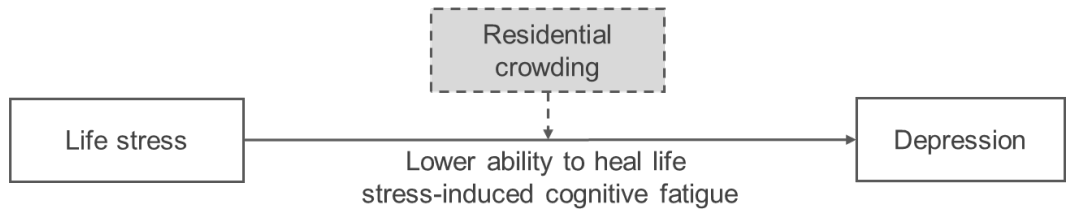

**Figure 1. Hypothesized mechanisms for the links between residential crowding, life stress, and depression.**

Alongside theoretical discussions, the residential crowding–mental well-being nexus is also supported by empirical evidence. According to a report by the U.S. Department of Housing and Development (HUD), residential crowding is usually measured by both average residential area per person and persons per room/bedroom (Blake et al., 2007). With respect to average residential area per person, using data from 13,367 participants in the 2010 wave of the China Family Panel Studies (or CFPS), a nationwide-representative survey, Hu and Coulter (2017) find that urban residents with higher living space per capita have better mental health. Focusing on China's migrants, Xie (2019) uses a 2008–2009 nationwide survey to find that on average, 10 more square meters of living space are associated with a 1.6 increase in GHQ-12 score[1]. Also looking at migrants, Li and Liu (2018) find that when controlling for perceived stress levels, the association between per-capita living space and mental well-being becomes statistically insignificant, implying that stress might be the mediator in the crowding-mental health association.

With respect to persons per room, Foye (2017) analyzes the British Household Panel Survey data and finds that individuals in circumstances with lower persons per room (i.e., less crowding) tend to have better mental health and life satisfaction. In a city-level study in Lahore, Pakistan, Khan et al. (2012) find that people living with more persons per have higher risks of depression and lower sense of control. Using a nationwide-representative sample in New Zealand, Pierse et al. (2016) find that those with higher "bedroom deficits" are more likely to experience psychological distress. However, if they focus on movers and control for household fixed effects, the association between residential crowding and mental health is only significant for those experiencing substantial crowding (i.e., a two-bedroom deficit) (Pierse et al., 2016).

The residential crowding–mental health association can differ between sociodemographic groups. For instance, Foye (2017) uses data from the UK to find that the residential crowding – mental health association is stronger for females than males, while the life satisfaction – mental health association is stronger for males than females. Although studies in China have not examined gender differences in the connections between residential crowding and mental health, they do find that when controlling for residential crowding levels, females tend to have worse mental health than males (Hu & Coulter, 2017; Xie, 2019). Other than gender differences, empirical evidence in China also shows that the association between living space and mental health is stronger for those with relatively higher socioeconomic status (Hu & Coulter, 2017). With respect to neighborhood context and housing type, Xie's (2019) study on China's migrants shows that the link between living space and mental health is stronger for those living in workers' dormitories than those occupying private rental housing units.

---

[1] GHQ-12 ranges from 0 to 36 and covers social functioning, depression and confidence. Higher scores indicate better mental health status.

Although existing theoretical reasoning and empirical evidence support the connection between residential crowding and mental health, there are still gaps in the literature. First, while studies quantitatively examining the residential crowding–mental health relationship have been emerging, research on the mechanisms of such relationship – specially the role stress plays – is still relatively rare. This study adopts the "stress process theory" originated in psychology (Pearlin & Bierman, 2013; Pearlin et al., 1981) and proposes a comprehensive framework to incorporate residential crowding, life stress and depression. Second, while residential crowding can be measured by both residential areas per person and number of persons per room, most studies only focus on one of the two. Focusing on both measures could help policy makers to better identify appropriate housing policy targets: whether to improve floor area or to reduce room deficits (Blake et al., 2007). Third, most studies covering Chinese cities focus on migrants, who certainly have been deprived through China's urbanization process. However, another group that has been "relatively deprived"—those with the urban *hukou* but still suffering residential crowding—also deserves attention due to their increasing contribution to urban inequality (Wang, 2004; Wu et al., 2010). This study aims to fill these gaps by examining the relationships among residential crowding, depression, and stress for Beijing's *hukou* holders, measuring residential crowding by both per-capita living space and persons per bedroom, and conducting various subsample analyses to identify subgroups that have higher "policy sensitivity" levels.

**Data and Methods**

*Data source and study sample*

The study sample comes from a survey conducted between November 2018 and April 2019 that covered seven of Beijing's 16 districts (Figure 2(a)). Among these seven districts, four are within the central-city area defined by the most recent master plan (in 2016): Xicheng is an Old-City district hosting various central-government agencies, Chaoyang has many financial and legal firms, Haidian is home of many universities and IT firms, and Fengtai is the wholesale and light industry center. The other three districts are suburban ones sitting outside the central-city area: Tongzhou was originally a suburban industrial hub and is being planned as the new headquarter of the municipal government, Changping is home of many large-scale edge-city communities, and Fangshan is mostly an exurban district with many recreational resorts (Feng et al., 2007; Sun, 2020).

The survey followed a multi-step stratified probability proportional to size sampling scheme. In each household, one main respondent was selected if the person (i) lived in Beijing, (ii) was 18–59 years old, and (iii) had formal Beijing residency (i.e., is a *hukou* holder). Participants responded to questions on residential crowding, life stress, property ownership, income, and other socioeconomic variables. The survey covered seven districts, 36 subdistricts, 168 communities, and 4,061 main respondents. The socioeconomic characteristics of these respondents are comparable to those of the 2015 1% Population Census – the most recent pre-covid nationwide survey[2] (National Bureau of Statistics of China, 2016).

---

[2] For the 4,061 survey respondents, the average age and share of females are 39.5 and 48.6%, respectively; for the 2015 Population Census, these two numbers are 39.9 and 50.0%, respectively.

The final study sample includes 1,613 main respondents who provided complete information on depression, residential crowding, life stress, and socioeconomic control variables. These 1,613 individuals were located in all seven districts, all 36 subdistricts, and 162 of the 168 communities in the survey, showing good spatial representativeness. The individuals excluded from the survey did not provide full information on one or more topics: 120 on depression; 1,618 on stress; 692 on residential crowding; and 18 on control variables. Figure 2(b) shows the spatial distribution of the 162 communities of the study sample.

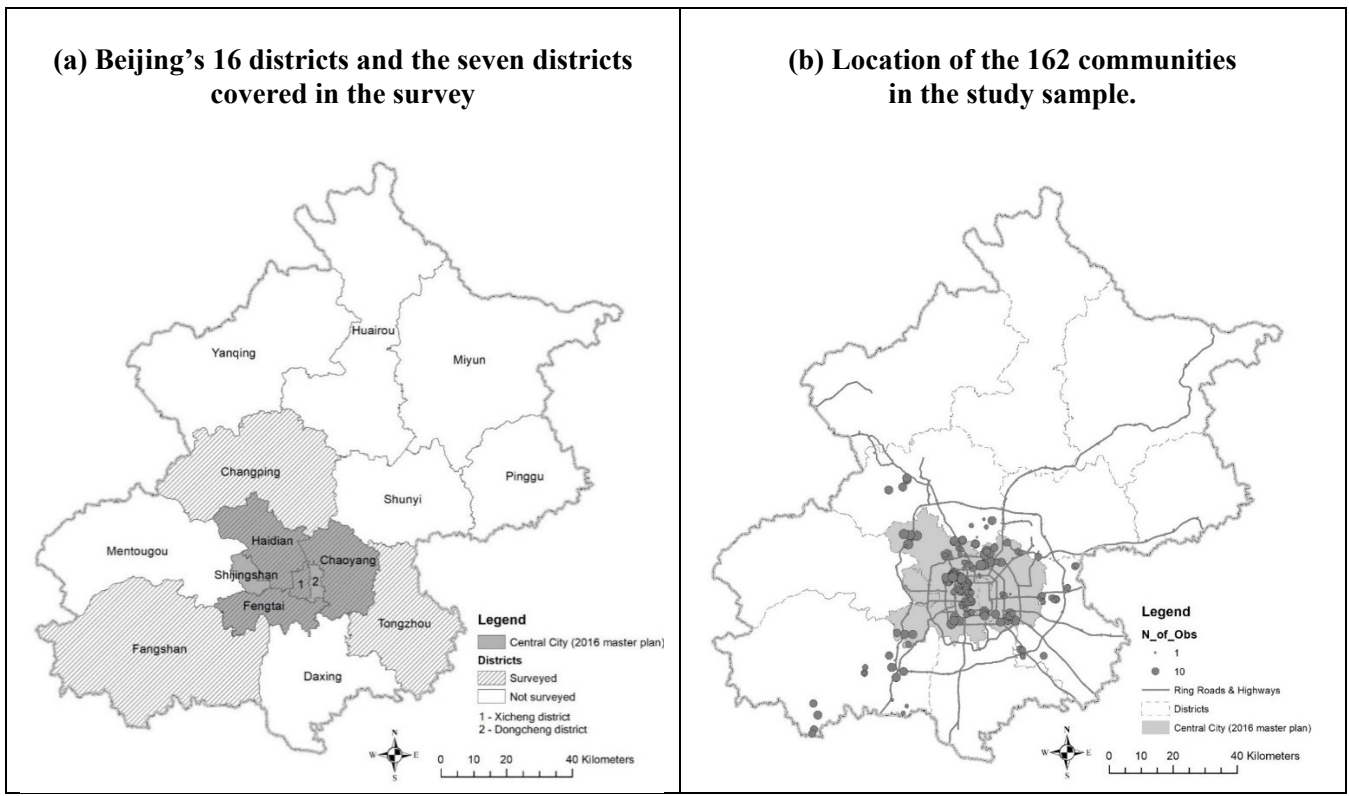

**Figure 2. Districts and communities covered in the study sample**
(Note: for (b), point sizes reflect the number of individuals in the community.)

*Depression, life stress, and residential crowding variables*

The outcome variable is depressive symptoms, measured by the Center for Epidemiological Studies' 10-item depression scale (CESD-10) (Andersen et al., 1994). This scale is frequently used in psychotherapy practice to screen for depression symptoms, and have been confirmed as a reliable screening scale across various cultural and geographic contexts (Bradley et al., 2010; Chen & Mui, 2014). The scale includes a list of 10 feelings[3], and surveys the respondent on the frequency in the past week with which they had each of the feelings. The response is then converted to a score ranging from 0 to 30, with higher values implying more depressed. We follow Andersen et al. (1994) and apply a cutoff score of 10. That is to say, a person is screened positive for depressive symptoms if their score is 10 or higher. Hence, we create a binary outcome variable that equals one if a respondent is screened positive for depression and zero if negative. We only include individuals who responded to all ten items.

Life stress is a continuous variable ranging from 0 to 15, with a higher value indicating a higher level of life stress. This variable is transformed from the survey questions on recent stress levels with respect to living costs, housing costs, child-rearing, work, and supporting parents. For each of these five stress types, the respondents choose from four categories: no stress, little stress, large stress, and very large stress, assigned values of 0, 1, 2, and 3, respectively. Hence, the final life stress variable is the aggregate of the five stress-type-specific values and ranges from 0 (least stress) to 15 (highest stress). Individuals missing any of the responses were dropped.

---

[3] The ten feelings are: (1) I was bothered by things that usually don't bother me; (2) I had trouble keeping my mind on what I was doing; (3) I felt depressed; (4) I felt that everything I did was an effort; (5) I felt hopeful about the future; (6) I felt fearful; (7) My sleep was restless; (8) I was happy; (9) I felt lonely; (10) I could not "get going".

Residential crowding is the study's key exposure variable. As noted above, we measure residential crowding by both square meters per person and persons per bedroom. Both measures are used extensively in academic research and policy discussions (UN-Habitat, 2010; United Nations Development Programme, 2010), while square meters per person being more commonly used in Chinese cities (Hu & Coulter, 2017; Li & Liu, 2018), and persons per room measures more common in the Western context (Foye, 2017; Pierse et al., 2016). Nevertheless, here we use both measures to capture different aspects of residential crowding, as square meters per person focuses on living space, and persons per bedroom emphasizes privacy. We decided to use persons per bedroom rather than persons per room, as a HUD report shows that the former has more reliable links with health outcomes (Blake et al., 2007). In the context of Chinese cities, "bedrooms" are normally referred to as "rooms," as real estate developers typically do not differentiate bedrooms from study rooms; it is up to residents to decide how they use their rooms. Hence, the "bedrooms" referred to in this study may also include study rooms or reading rooms. Finally, as there are no widely accepted cutoff points for persons per bedroom in the Chinese context, we use both 1 and 1.5 as cutoff points and propose a three-level categorical variable: up to 1 person per bedroom, 1.01–1.50 persons per bedroom, and 1.51 or more persons per bedroom.

*Control variables*

The control variables include eleven sociodemographic characteristics, as well as district fixed effects. The sociodemographic variables aim to control for potential confounding factors, including gender (male or female), age (in 2018), living with a spouse or partner, living with parents or grandparents, living with children or grandchildren, employment status, annual household income (in 10,000 Chinese *yuan* or 1,546 U.S. Dollars), owning residential property in Beijing (disregarding ownership status of the current residence; a household wealth indicator[4]), living in condominium (with "no" indicating living in non-market housing, such as *hutong*, *danwei*, resettled housing, or villages), level of education (a categorical variable of lower than high school, high school/some college, and college degree or higher), job type (a categorical variable of government, managerial positions, professions such as engineering or law, agriculture, manufacturing/construction, and others). District fixed effects is a categorical variable for the districts in which the respondents reside (Xicheng, Chaoyang, Fengtai, Haidian, Fangshan, Tongzhou and Changping). District fixed effects can control for district-specific factors associated with both living space and mental health, such as the socio-economic stereotypes.

*Model specifications*

To examine the relationships among residential crowding, life stress, and depression, we propose the models following equation (1):

$$Depress_i = \beta_0 + \beta_1 Crowding_i + \beta_2 LifeStress_i + \beta_3 SocioDem_i + \beta_4 DistrictFE_i , \quad (1)$$

where $Depress_i$ is a dummy variable which equals 1 if individual $i$ is screened positive for depression, $Crowding_i$ measures levels of residential crowding, and can be either square meters per person or persons per bedroom, $LifeStress_i$ measures life stress levels, and $SocioDem_i$ and $DistrictFE_i$ refers to a set of socio-demographic and district fixed effects control variables.

---

[4] This "residential property ownership" variable is mainly a wealth indicator. Residential property is a major asset for China's urban residents, and has been a major contributor to wealth inequality (He et al., 2020; P. Zhang et al., 2021).

To further test for the potential mechanisms of the relationships among residential crowding, life stress, and depression, we propose two additional models:

$$Depress_i = \beta_0 + \beta_1\, Crowding_i + \beta_2\, SocioDem_i + \beta_3\, DistrictFE_i, \quad (2)$$

$$Depress_i = \beta_0 + \beta_1\, Crowding_i + \beta_2\, LifeStress_i + \beta_3\, Crowding_i \times LifeStress_i \\ + \beta_4\, SocioDem_i + \beta_5\, DistrictFE_i. \quad (3)$$

If residential crowding is associated with depression through increased life stress (i.e., life stress is the mediator; see hypothesis b in Figure 1), $LifeStress_i$ would have a statistically significant mediation effect which can be identified using the Sobel test; otherwise, residential crowding is associated with depression through increased residential-specific stress (hypothesis a in Figure 1) (Baron & Kenny, 1986; Sobel, 1982). Intuitively, if a mediation effect exists, the coefficients of $Crowding_i$ in equations (1) and (2) would also differ substantially. If the coefficient of $Crowding_i \times LifeStress_i$ in equation (3) is statistically significant, then it shows that levels of residential crowding can make people more prone to depression from life stress (i.e., life stress is the moderator; see hypothesis c in Figure 1) (Baron & Kenny, 1986).

All models are logit regressions, as they all have binary dependent variables. Since the value of the dependent variable is imbalanced (5.5% vs. 94.5%), estimations using regular logit models may be biased. Hence, we follow the literature and apply Firth's corrections for rare events in the regressions (King & Zeng, 2001; Leitgöb, 2013).

**Results**

*Descriptive statistics*

Table 1 presents the descriptive statistics for the study sample, showing that 5.5%

screened positive for depressive symptoms. To our knowledge, there have been no other citywide surveys in Beijing covering depressive symptoms or using CESD-10. We compared our study sample with the 2018 wave of the China Family Panel Studies (CFPS), which surveyed eight of the ten CESD-10 items (Peking University ISSS, 2020). Based on the CFPS, the average eight-item CESD score for the Beijing *hukou* holders is 3.76, while the same metric for our study sample is 4.21. For residential crowding, the average per-capita living space is 36.8 square meters, which is slightly higher than the city-level statistics in 2015 (31.7 square meters), the most recent publicly available statistics (Beijing Municipal Bureau of Statistics, 2016). The difference might be due to the fact that the 2015 city-level statistic includes both locals and migrants, while our sample only includes *hukou* holders. Regarding persons per bedroom, 37% of the study sample lives in a household with one or fewer persons per bedroom, and 13.4% of them live in a household with higher than 1.5 persons per bedroom. The average stress score of the study sample is 5.8 out of 15. In addition, 66% of the study sample live in condominiums, which means that 34% live in non-market properties, such as *hutong* (Old City heritage blocks), *danwei* (state-owned employer-provided housing), resettled housing (after land acquisitions), or villages. Sub-sample statistics show that the residential crowding situation is relatively worse among males, those living with parents, those living with children, and condominium residents.

Table 1. Descriptive statistics for the study sample (N = 1,613)

|  | Mean or % | SD | Min | Max |
|---|---|---|---|---|
| Depression (y/n) | 0.055 | 0.227 | 0 | 1 |
| Square meters per person | 36.792 | 27.929 | 5 | 200 |
| Persons per bedroom (%) |  |  |  |  |
| *1-or-fewer persons per bedroom* | 36.9% |  |  |  |
| *1.01~1.50 persons per bedroom* | 49.7% |  |  |  |
| *1.51-or-more persons per bedroom* | 13.4% |  |  |  |
| Stress score (0-15) | 5.817 | 3.738 | 0 | 15 |
| Female | 0.509 | 0.500 | 0 | 1 |
| Age in 2018 (years) | 41.484 | 11.395 | 18 | 59 |
| Living with spouse/partner | 0.761 | 0.427 | 0 | 1 |
| Living with parents/grandparents | 0.403 | 0.491 | 0 | 1 |
| Living with children/grandchildren | 0.570 | 0.495 | 0 | 1 |
| Currently employed | 0.631 | 0.483 | 0 | 1 |
| Household income in 2017 (in 10k) | 20.067 | 39.555 | 0 | 1000 |
| Owning residential property in Beijing | 0.660 | 0.474 | 0 | 1 |
| Condominium | 0.536 | 0.499 | 0 | 1 |
| Job type |  |  |  |  |
| *Government* | 14.0% |  |  |  |
| *Manager* | 6.7% |  |  |  |
| *Specialist* | 15.5% |  |  |  |
| *Service* | 14.3% |  |  |  |
| *Agriculture* | 2.7% |  |  |  |
| *Manufacturing/construction* | 8.8% |  |  |  |
| *Other* | 38.0% |  |  |  |
| District |  |  |  |  |
| *Xicheng* | 13.0% |  |  |  |
| *Chaoyang* | 18.3% |  |  |  |
| *Fengtai* | 12.4% |  |  |  |
| *Haidian* | 30.1% |  |  |  |
| *Fangshan* | 8.4% |  |  |  |
| *Tongzhou* | 8.3% |  |  |  |
| *Changping* | 9.5% |  |  |  |

*Residential crowding, life stress, and depression*

The regression models in Table 2 show that a higher level of residential crowding is associated with a higher risk of depression. Drawing upon the baseline models in Columns 2 and 5, when holding continuous variables at their means and categorical variables at their modes, every 10 additional square meters per person is associated with a 27.7% lower probability of depression, and individuals with more than 1.5 persons per bedroom are 1.2 times more likely to have depressive symptoms than those with one or fewer persons per bedroom. To test for the mechanisms of the relationships among residential crowding, life stress, and mental health, we first conduct the Sobel test for mediation effects, and find that the residential crowding–depression association is not mediated by life stress (results not shown). Intuitively, when comparing the coefficients of the residential crowding variables in Columns 1 vs. 2 and Columns 4 vs. 5 in Table 2, the coefficients of area per person and persons per bedroom have very small differences with and without controlling for life stress. Secondly, we examine the coefficients of the residential crowding–life stress interaction variables in Columns 3 and 6 and find that neither is statistically significant. Such findings indicate that residential crowding does not moderate the association between life stress and depression. In sum, among the different hypotheses posited in Figure 1, it is most likely that residential crowding is associated with mental health through residential space-specific stress.

**Table 2. Depression, residential crowding, and life stress**

|  | (1) Area per person, no stress | (2) Area per person, w/ stress | (3) Area per person, interact w/ stress | (4) Persons per bedroom no stress | (5) Persons per bedroom w/ stress | (6) Persons per bedroom, interact w/ stress |
|---|---|---|---|---|---|---|
| Square meters per person | -0.029*** | -0.028*** | -0.018 | | | |
|  | [0.008] | [0.008] | [0.012] | | | |
| Persons per bedroom | | | | | | |
| *1-or-fewer persons per bedroom* | | | | (ref.) | (ref.) | (ref.) |
| *1.01~1.50 persons per bedroom* | | | | 0.682** | 0.686** | 0.895* |
|  | | | | [0.285] | [0.285] | [0.544] |
| *1.51-or-more persons per bedroom* | | | | 1.186*** | 1.198*** | 0.866 |
|  | | | | [0.372] | [0.376] | [0.702] |
| Stress score (0-15) | | 0.113*** | 0.155** | | 0.118*** | 0.123* |
|  | | [0.040] | [0.064] | | [0.040] | [0.065] |
| Square meters per person # Stress score | | | -0.002 | | | |
|  | | | [0.002] | | | |
| Persons per bedroom # Stress score | | | | | | |
| *1-or-fewer ppl per bedroom # stress* | | | | | | (ref.) |
| *1.01~1.50 persons per bedroom # stress* | | | | | | -0.030 |
|  | | | | | | [0.072] |
| *1.51-or-more persons per bedroom # stress* | | | | | | 0.046 |
|  | | | | | | [0.086] |
| Female | -0.007 | 0.015 | 0.004 | -0.077 | -0.062 | -0.070 |
|  | [0.250] | [0.252] | [0.252] | [0.252] | [0.254] | [0.254] |
| Age in 2018 (years) | 0.032** | 0.037** | 0.035** | 0.033** | 0.039*** | 0.040*** |
|  | [0.015] | [0.015] | [0.016] | [0.015] | [0.015] | [0.015] |
| Living with spouse/partner | -1.647*** | -1.736*** | -1.693*** | -1.673*** | -1.783*** | -1.761*** |
|  | [0.412] | [0.407] | [0.409] | [0.412] | [0.408] | [0.411] |
| Living with parents/grandparents | -0.496 | -0.450 | -0.443 | -0.623* | -0.587* | -0.602* |
|  | [0.322] | [0.321] | [0.320] | [0.324] | [0.324] | [0.325] |
| Living with children/grandchildren | 0.043 | -0.264 | -0.257 | 0.041 | -0.299 | -0.327 |
|  | [0.297] | [0.312] | [0.313] | [0.305] | [0.321] | [0.324] |
| Currently employed | -0.029 | -0.256 | -0.221 | 0.057 | -0.185 | -0.169 |
|  | [0.938] | [0.940] | [0.943] | [0.933] | [0.936] | [0.935] |
| Natural log of household income (in 10k) in 2017 | 0.779*** | 0.879*** | 0.860*** | 0.700*** | 0.807*** | 0.810*** |
|  | [0.203] | [0.206] | [0.203] | [0.196] | [0.199] | [0.200] |
| Owning residential property in Beijing | 0.128 | 0.200 | 0.216 | 0.082 | 0.149 | 0.167 |
|  | [0.324] | [0.325] | [0.325] | [0.312] | [0.313] | [0.315] |
| Living in condominium | -1.210*** | -1.298*** | -1.311*** | -1.065*** | -1.171*** | -1.163*** |
|  | [0.269] | [0.271] | [0.272] | [0.271] | [0.274] | [0.274] |
| Education | | | | | | |
| *Less than High School* | (Ref.) | (Ref.) | (Ref.) | (Ref.) | (Ref.) | (Ref.) |
| *High school/some college* | -0.287 | -0.348 | -0.350 | -0.164 | -0.209 | -0.208 |
|  | [0.341] | [0.344] | [0.344] | [0.339] | [0.343] | [0.342] |
| *College or higher* | -0.927* | -0.949* | -0.940* | -0.720 | -0.728 | -0.772 |
|  | [0.486] | [0.489] | [0.488] | [0.483] | [0.486] | [0.489] |
| Job type | | | | | | |
| *Government* | (Ref.) | (Ref.) | (Ref.) | (Ref.) | (Ref.) | (Ref.) |
| *Manager* | 1.030** | 1.092** | 1.118** | 1.023** | 1.073** | 1.069** |
|  | [0.520] | [0.527] | [0.529] | [0.518] | [0.525] | [0.526] |
| *Specialist* | -0.371 | -0.219 | -0.204 | -0.440 | -0.280 | -0.252 |
|  | [0.593] | [0.603] | [0.604] | [0.593] | [0.604] | [0.605] |
| *Service* | 0.841* | 0.955** | 0.972** | 0.822* | 0.946** | 0.943** |
|  | [0.453] | [0.459] | [0.461] | [0.450] | [0.456] | [0.456] |
| *Agriculture* | 0.685 | 0.974 | 0.996 | 0.320 | 0.601 | 0.610 |
|  | [0.805] | [0.810] | [0.811] | [0.790] | [0.798] | [0.797] |
| *Manufacture/construction* | 0.281 | 0.431 | 0.452 | 0.246 | 0.400 | 0.396 |
|  | [0.532] | [0.538] | [0.540] | [0.527] | [0.533] | [0.535] |
| *Other* | 0.216 | 0.409 | 0.438 | 0.181 | 0.367 | 0.366 |
|  | [0.985] | [0.988] | [0.991] | [0.980] | [0.984] | [0.983] |
| District fixed effects |  |  |  |  |  |  |
| *Xicheng* | (Ref.) | (Ref.) | (Ref.) | (Ref.) | (Ref.) | (Ref.) |
| *Chaoyang* | 2.477*** | 2.299*** | 2.304*** | 2.068** | 1.924** | 1.981** |
|  | [0.883] | [0.882] | [0.880] | [0.876] | [0.874] | [0.877] |
| *Fengtai* | 2.622*** | 2.413*** | 2.375*** | 2.322** | 2.134** | 2.163** |
|  | [0.914] | [0.919] | [0.919] | [0.909] | [0.911] | [0.913] |
| *Haidian* | 2.463*** | 2.202** | 2.190** | 2.243*** | 2.006** | 2.058** |
|  | [0.856] | [0.859] | [0.858] | [0.850] | [0.853] | [0.855] |
| *Fangshan* | 3.021*** | 2.509*** | 2.526*** | 2.775*** | 2.304** | 2.341** |
|  | [0.942] | [0.953] | [0.953] | [0.927] | [0.938] | [0.939] |
| *Tongzhou* | 3.757*** | 3.446*** | 3.425*** | 3.403*** | 3.106*** | 3.144*** |
|  | [0.881] | [0.885] | [0.883] | [0.872] | [0.874] | [0.876] |
| *Changping* | 2.915*** | 2.672*** | 2.642*** | 2.684*** | 2.465*** | 2.501*** |
|  | [0.909] | [0.913] | [0.911] | [0.900] | [0.903] | [0.904] |
| Constant | -5.993*** | -6.726*** | -6.908*** | -7.108*** | -7.888*** | -8.001*** |
|  | [1.655] | [1.699] | [1.704] | [1.619] | [1.668] | [1.709] |
| Number of observations | 1613 | 1613 | 1613 | 1613 | 1613 | 1613 |

Note: Logit regression with Firth's penalized maximum likelihood methods; the dependent variable is screening positive for depression. *,**,*** indicate significance at 0.10, 0.05 and 0.01 levels. Standard errors are in parentheses.

*Effects by gender*

Table 3 shows that the association between residential crowding and depression is stronger for females than males. As Columns 2 and 4 indicate, the coefficients of residential crowding variables—measured by both square meters per person and persons per bedroom—are significant for females; in contrast, Columns 3 and 6 show that neither residential crowding variable is significantly associated with depression for males. Additionally, the magnitude of the residential crowding coefficients is larger for females than for males.

**Table 3. Depression and residential crowding: by gender**

|  | (1) All | (2) Female | (3) Male | (4) All | (5) Female | (6) Male |
|---|---|---|---|---|---|---|
| Square meters per person | -0.028*** | -0.029*** | -0.021* |  |  |  |
|  | [0.008] | [0.010] | [0.011] |  |  |  |
| Persons per bedroom |  |  |  |  |  |  |
|   *1-or-fewer persons per bedroom* |  |  |  | (ref.) | (ref.) | (ref.) |
|   *1.01~1.50 persons per bedroom* |  |  |  | 0.686** | 0.799* | 0.413 |
|  |  |  |  | [0.285] | [0.431] | [0.391] |
|   *1.51-or-more persons per bedroom* |  |  |  | 1.198*** | 1.391*** | 0.934 |
|  |  |  |  | [0.376] | [0.516] | [0.573] |
| Stress score (0-15) | 0.113*** | 0.165*** | 0.077 | 0.118*** | 0.167*** | 0.086 |
|  | [0.040] | [0.056] | [0.057] | [0.040] | [0.056] | [0.057] |
| Controls | Yes | Yes | Yes | Yes | Yes | Yes |
| District fixed effects | Yes | Yes | Yes | Yes | Yes | Yes |
| Observations | 1613 | 821 | 792 | 1613 | 821 | 792 |

Note: Logit regressions with Firth's penalized maximum likelihood methods, following Equation (1) in the "Model specifications" section. The dependent variable is screening positive for depression. Control variables include age, living with spouse, living with parents, living with children, employed, natural log of household income, owning residential property in Beijing, living in condominium, education and job type. *, **, *** indicate significance at 0.10, 0.05 and 0.01 levels. Standard errors are in parentheses.

*Effects by household structure*

Tables 4 includes the subsample analysis based on house structure, and show that the association between residential crowding and depression is stronger for those living with children/grandchildren as well as those not living with parents/grandparents. Table 4(A) shows that the association is statistically significant for those living with children (Columns 2 and 5) and not significant for those not living with children (Columns 3 and 6). Table 4(A) also shows that the magnitudes of the coefficients of the two residential crowding variables are larger for those living with children than those not living with children. With respect to living with parents, Table 4(B) shows that both residential crowding variables are significantly associated with depression for those not living with parents (Columns 9 and 12), and neither variable is significantly associated with depression for those living with parents (Columns 8 and 11). Similarly, Table 4(B) shows that the magnitudes of the coefficients of the residential crowding variables are larger for those not living with parents than for those living with parents.

Table 4. Depression and residential crowding: by household structure

| Panel A – By living with children | (1) All | (2) Living w/ children | (3) Not living w/ children | (4) All | (5) Living w/ children | (6) Not living w/ children |
|---|---|---|---|---|---|---|
| Square meters per person | -0.028*** | -0.035*** | -0.020* | | | |
|  | [0.008] | [0.012] | [0.010] | | | |
| Persons per bedroom | | | | | | |
| *1-or-fewer persons per bedroom* | | | | (ref.) | (ref.) | (ref.) |
| *1.01~1.50 persons per bedroom* | | | | 0.686** | 0.762* | 0.323 |
|  | | | | [0.285] | [0.410] | [0.451] |
| *1.51-or-more persons per bedroom* | | | | 1.198*** | 1.280** | 0.576 |
|  | | | | [0.376] | [0.497] | [0.643] |
| Stress score (0-15) | 0.113*** | 0.104** | 0.172** | 0.118*** | 0.114** | 0.189*** |
|  | [0.040] | [0.053] | [0.074] | [0.040] | [0.053] | [0.072] |
| Controls | Yes | Yes | Yes | Yes | Yes | Yes |
| District fixed effects | Yes | Yes | Yes | Yes | Yes | Yes |
| Number of observations | 1613 | 919 | 694 | 1613 | 919 | 694 |

| Panel B – By living with parents | (7) All | (8) Living w/ parents | (9) Not living w/ parents | (10) All | (11) Living w/ parents | (12) Not living w/ parents |
|---|---|---|---|---|---|---|
| Square meters per person | -0.028*** | -0.025* | -0.029*** | | | |
|  | [0.008] | [0.015] | [0.010] | | | |
| Persons per bedroom | | | | | | |
| *1-or-fewer persons per bedroom* | | | | (ref.) | (ref.) | (ref.) |
| *1.01~1.50 persons per bedroom* | | | | 0.686** | -0.110 | 1.361*** |
|  | | | | [0.285] | [0.528] | [0.388] |
| *1.51-or-more persons per bedroom* | | | | 1.198*** | 1.069* | 0.984* |
|  | | | | [0.376] | [0.607] | [0.541] |
| Stress score (0-15) | 0.113*** | 0.012 | 0.160*** | 0.118*** | 0.019 | 0.169*** |
|  | [0.040] | [0.075] | [0.053] | [0.040] | [0.075] | [0.052] |
| Controls | Yes | Yes | Yes | Yes | Yes | Yes |
| District fixed effects | Yes | Yes | Yes | Yes | Yes | Yes |
| Number of observations | 1613 | 650 | 963 | 1613 | 650 | 963 |

Note: Logit regressions with Firth's penalized maximum likelihood methods, following Equation (1) in the "Model specifications" section. The dependent variable is screening positive for depression. Control variables include gender, age, living with spouse, living with parents, living with children, employed, natural log of household income, owning residential property in Beijing, living in condominium, education and job type. *, **, *** indicate significance at 0.10, 0.05 and 0.01 levels. Standard errors are in parentheses.

*Effects by housing and neighborhood types*

Table 5 examines the residential crowding–depression association by different housing and neighborhood types. Here, we divide the study sample into two housing types: condominium and non-condominium. Condominium refers to housing developed and sold by private developers, and non-condo refers to housing obtained through non-market channels, including *hutong* (Old City heritage blocks), *danwei* (state-owned employer-provided housing), resettled housing (after land acquisitions), and villages. In China's context, condo residents are more likely to obtain their housing voluntarily through the open market, and non-condo residents are more likely to obtain their housing involuntarily through inheritance, employment, or resettlement. Columns 3 and 6 in Table 5 show that residential crowding is significantly associated with depression for non-condo dwellers, while Columns 2 and 5 show that the residential crowding–depression association is not significant for condo dwellers. Besides, the coefficients for the non-condo subsample are greater in magnitude than those for the condo subsample. In sum, the association between residential crowding and depression is stronger for non-condo residents than for condo residents.

Table 5. Depression and residential crowding: by neighborhood type

| | (1) All | (2) Condo | (3) Non-condo | (4) All | (5) Condo | (6) Non-condo |
|---|---|---|---|---|---|---|
| Square meters per person | -0.028*** | -0.023 | -0.026*** | | | |
| | [0.008] | [0.020] | [0.008] | | | |
| Persons per bedroom | | | | | | |
|   *1-or-fewer persons per bedroom* | | | | (ref.) | (ref.) | (ref.) |
|   *1.01~1.50 persons per bedroom* | | | | 0.686** | -0.293 | 0.978*** |
| | | | | [0.285] | [0.506] | [0.349] |
|   *1.51-or-more persons per bedroom* | | | | 1.198*** | 0.205 | 1.368*** |
| | | | | [0.376] | [0.629] | [0.505] |
| Stress score (0-15) | 0.113*** | 0.088 | 0.101* | 0.118*** | 0.099 | 0.105* |
| | [0.040] | [0.061] | [0.058] | [0.040] | [0.060] | [0.058] |
| Controls | Yes | Yes | Yes | Yes | Yes | Yes |
| District fixed effects | Yes | Yes | Yes | Yes | Yes | Yes |
| Number of observations | 1613 | 864 | 749 | 1613 | 864 | 749 |

Note: Logit regressions with Firth's penalized maximum likelihood methods, following Equation (1) in the "Model specifications" section. The dependent variable is screening positive for depression. Control variables include gender, age, living with spouse, living with parents, living with children, employed, natural log of household income, owning residential property in Beijing, education and job type. *, **, *** indicate significance at 0.10, 0.05 and 0.01 levels. Standard errors are in parentheses.

*Robustness checks*

    We conducted a few sensitivity analyses to assess the robustness of the findings. First, we followed Andersen et al. (1994) and expanded the study sample by including individuals with one missing CESD-10 item. Models with this expanded sample yielded similar results. Second, to test for the sensitivity of the 10-score cutoff point, we ran our full-sample models again with cutoff scores of 9 and 11. The signs and significance of both residential crowding and life stress variables remain unchanged. Third, we replaced the binary outcome variable with the raw CESD-10 score (0-30) and ran negative binomial models. Regression models using the raw score still support our main conclusions. Fourth, to evaluate the sensitivity of the life stress measurements, we changed the continuous (0–15) life stress scores into a dummy stress variable, with 1 indicating the stress level is above the median (6 out of 15) and 0 otherwise. The full-

sample models using this dummy life stress variable yield similar results to those with continuous life stress scales. Fifth, we reran the main models with three additional control variables: door-to-door commute time, population density (800m radius), and greenspace accessibility within 800m, adding these variables does not impact the main findings. Finally, we estimated the variance inflation factors (VIF) of the full models and did not find evidence of multicollinearity.

**Discussion**

Using survey data from Beijing, China, we find that residential crowding—measured by both square meters per person and persons per bedroom—is associated with a higher probability of depression. We propose three hypotheses for the potential mechanisms that connect residential crowding with depression, and find that residential crowding is associated with depression through increased residential space-specific stress rather than increased life stress; additionally, we do not find evidence of residential crowding moderating the life stress–depression association. Moreover, we find that the residential crowding–depression association is relatively stronger for females, those living with children/grandchildren, those not living with parents/grandparents, and those not living in condominiums.

The finding that square meters per person and persons per bedroom are associated with a higher propensity for depression is in consonance with previous studies on residential crowding and depression in Chinese and western cities (Foye, 2017; Hu & Coulter, 2017; Li & Liu, 2018; Pierse et al., 2016; Xie, 2019). As noted earlier, to measure residential crowding levels, most studies on Asian cities use square meters per person, and most studies on North American and European cities use persons per room measures. By using both measures, this study shows the robustness of the residential crowding–depression association. We also find that residential crowding is associated with depression through increased residential space-specific stress rather than increased life stress. This finding joins the work of Li and Liu (2018) to extend the theoretical discussions on this topic by exploring the role stress plays in the residential crowding–depression relationship. Additionally, the coefficients of district fixed effects show that residents in central-city districts are relatively less depressed, which likely reflects the district-level difference in commute time, built environment, industrial specializations and social stereotypes (Feng et al., 2007; Sun, 2020).

We find that females are relatively more "mentally sensitive" to crowded living space than males. This finding supports the theoretical discussion in psychology that females are relatively more vulnerable to stress and more likely to be depressed due to their affective, biological, and cognitive vulnerabilities (Hyde et al., 2008). Empirically, this finding resonates with Foye's (2017) on the same question in the UK context and also connects with the findings of Hu and Coulter (2017) and Li and Liu (2018) that females are more likely to be depressed than males in Chinese cities. Admittedly, the coefficient of the gender variable is not statistically significant in the full-sample baseline models; nevertheless, the female respondents in our study sample do have a higher rate of depression than the males (5.6% vs. 5.3%), without controlling for other covariates. The insignificance of the gender variable in the full-sample models might be due to confounding between the gender variable and other control variables, such as living with a partner, parents, and children.

Our findings that those not living with parents and those living with children have a relatively stronger association between residential crowding and depression imply that: (i) living with parents can make people more mentally robust against external stresses and (ii) living with children might make people more mentally vulnerable due to childcare duties. These findings extend existing studies on the impact of intergenerational co-living on mental health (Brunello & Rocco, 2019; Courtin & Avendano, 2016), using the lens of housing and residential crowding to explore relative "mental vulnerability to crowding" for different household structures.

Finally, we find that the residential crowding–depression association is stronger in non-condo communities than in condo communities. To our knowledge, no prior studies have specifically focused on different "effect sizes" across different neighborhood types in Chinese cities. Here, we hypothesize two possible explanations. First, this association may be due to the fact that condominiums are obtained through the open market. Hence, those living in condos are more likely to have obtained their homes by choice. In contrast, those living in non-condo communities are more likely to have obtained their residences through inheritance, employment, or resettlement rather than by choice. Second, it is also possible that condo communities have certain built environment or social environment characteristics that can help to buffer against the negative mental health impacts of residential crowding. We acknowledge that the non-condo communities in this study include a diverse set of non-market housing types that range from Old City traditional blocks (*hutong*) to ex-urban villages, and we are unable to further divide them into more detailed housing types due to sample size limitations. Nevertheless, further discussion of both hypotheses requires additional data and is beyond the scope of this research.

Our findings highlight the importance of living space equality for promoting mental health and subjective well-being of residents in big cities. Specifically, both rooms and floor space matter. Although Beijing's "registered residents" (i.e., *hukou* holders) enjoy larger living spaces than migrants on average, the unequal distribution of living space among these *hukou* holders has important policy implications (Huang & Jiang, 2009). Those who have been left behind in China's recent rapidly-growing real estate development process deserve attention from planners and policy makers. Public housing programs that aim to provide sufficient floor area and adequate number of rooms for lower-income urban *hukou* holders would not only ensure equality of living space among urban residents, but also improve these residents' mental health status. For example, the "joint-ownership housing program" (*gong you chan quan zhu fang*) provides for-sale subsidized flats for non-condo owners with limitations on resales in the open real estate market, making it a good option for the lower-income households that need to improve their living space but do not intend to engage in real estate investments. In addition, the "public rental housing program" (*gong zu fang*) provides for-rent subsidized flats for non-condo owners, making it a good option for lower-income households who wish to expand their living space and with no intention/ability to own a flat. For the households owning residential properties but still suffer from residential crowding, the government should work with the financial sector to provide them adequate support to upgrade their living conditions in the open real estate market.

Although lottery should be the default method of allocating subsidized housing units, prioritizing applicants who need living space more urgently may be advisable. Our findings highlight a few subgroups that are relatively more "mentally sensitive" to cramped living space: females, young parents with children, those living without parents, and non-condo dwellers. Applicants with these characteristics may deserve prioritization on waitlists or higher chances in lotteries. One example of such policy is Singapore's public housing system, which has many schemes that prioritize housing allocation for young parents, young family living close to their parents, and first-time applicants (Centre for Livable Cities Singapore & Shanghai Municipal Commission of Housing Urban-Rural Development and Management, 2020). Although China's big cities differ from Singapore in spatial, socioeconomic, and institutional contexts, some of these policy practices may still be transferrable.

This study has the following limitations, which should all motivate future research. First, the study sample is a cross-sectional dataset, and the relationships between residential crowding and depression can only be interpreted as associations rather than causations; future studies could utilize longitudinal data and try to build causal relationships. Second, as the study focuses on *hukou* holders, policy implications drawn from this paper cannot be extended to migrants. Third, this study is unable to examine the residential crowding-depression associations for each specific non-condo neighborhood type due to limited sample size. Future studies specifically focusing on Old City traditional blocks *(hutong)* or resettlement communities would greatly extend the literature. Fourth, although the survey respondents in and outside the study sample have similar geographical distributions, they may differ in other non-spatial characteristics. While controlling for a comprehensive list of control variables can adjust for the bias created by these variables (King et al., 1994), we are not able to rule out the possibility that those in and outside the sample differ in unobserved ways (e.g., personalities or beliefs).

**Conclusions**

Using survey data for 1,613 residents in Beijing, China, this study has found that residential crowding—measured by both square meters per person and persons per bedroom—is associated with a higher risk of depression. We also examined the mechanisms of the associations between residential crowding and depression by testing the role stress plays in this relationship, and we found that living in a crowded house is associated with depression through increased residential-specific stress rather than increased life stress. The findings highlight an important group of urban residents who deserve attention from housing policy makers: long-term urban residents left behind by the rapid development of China's real estate markets who have limited living space. Among these "left-behind residents," females, those living with children, those not living with parents, and those living in non-condo communities are more "mentally vulnerable" to crowded housing. Public housing programs aiming to provide living space for these left-behind residents—who normally have low incomes and do not own condominiums—not only would increase living space equality but also could improve the mental health status and subjective well-being levels of urban dwellers.